# An Empirical Review of Adversarial Defenses

## Ayush Goel

## Abstract

From face recognition systems installed in phones to self-driving cars, the field of AI is witnessing rapid transformations and is being integrated into our everyday lives at an incredible pace. Any major failure in these system's predictions could be devastating, leaking sensitive information or even costing lives (as in the case of self-driving cars). However, deep neural networks, which form the basis of such systems, are highly susceptible to a specific type of attack, called adversarial attacks. A hacker can, even with bare minimum computation, generate adversarial examples (images or data points that belong to another class, but consistently fool the model to get misclassified as genuine) and crumble the basis of such algorithms. In this paper, we compile and test numerous approaches to defend against such adversarial attacks. Out of the ones explored, we found two effective techniques, namely Dropout and Denoising Autoencoders, and show their success in preventing such attacks from fooling the model. We demonstrate that these techniques are also resistant to both higher noise levels as well as different kinds of adversarial attacks (although not tested against all). We also develop a framework for deciding the suitable defense technique to use against attacks, based on the nature of the application and resource constraints of the Deep Neural Network.

## 1. Introduction

### 1.1 Context

The generous availability of fast computing power and storage at lower costs, coupled with a rapid proliferation of connected devices, has led to a drastic increase in the volume, velocity, and variety of data being generated. This big data revolution has heralded a new wave of Machine Learning techniques called Deep Learning (performance comparison shown in Fig 1), which focuses on using Deep Neural Networks to perform complex tasks such as Real-Time Object Detection, and even Cancer Detection, sometimes achieving a similar performance as human experts in the field.

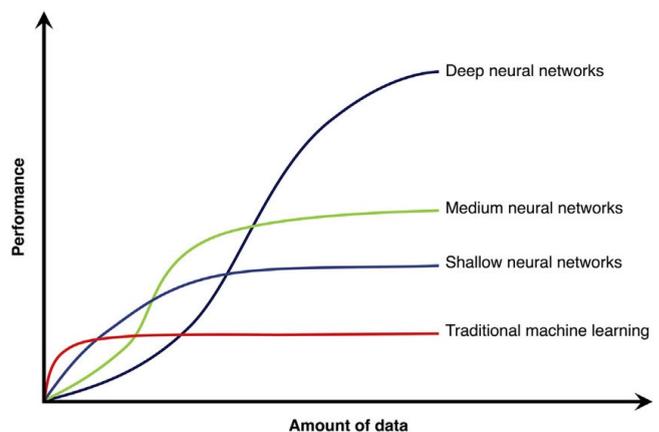

Fig 1 - Performance of Different Techniques vs Size of Data

Due to their high performance, Deep Neural Networks are being readily deployed on numerous tasks across diverse domains, from healthcare and marketing to more sensitive areas such as security and safe transportation. Although these models were originally thought to be robust to a wide range of inputs, a recent paper by Ian Goodfellow, et al. showed that such Neural Networks are, in practice, highly vulnerable to finely tuned input data. In their paper, the authors demonstrated that by carefully probing the Neural Network, and modifying the input image as required, they could generate "adversarial examples", images or data points that belong to another class, but consistently fool the model to get misclassified as genuine.

Furthermore, their paper showed that generating such images didn't even require access to the original Deep Neural Network. Even a smaller Network, trained to perform a similar task could be probed, and with a high probability, the same image would fool the deeper inaccessible model as well. This has severe implications since attackers don't need to break any form of encryption to fool important Deep Neural Networks such as ones for Face Verification. Instead, they need only large amounts of data, which is generally publicly accessible.

Over time, more research has been done on the field, and researchers have discovered numerous methodologies for orchestrating such attacks, some even faster than the original method. However, this research has not received the attention it should have. Due to this, many commercially deployed networks are vulnerable to such attacks, causing significant risk to society.

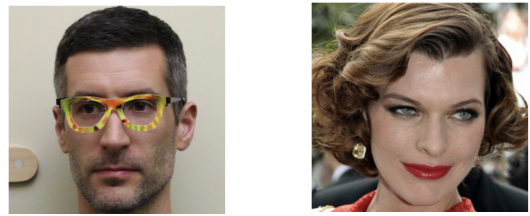

Input Image          Face Recognized by NN

Fig 2 - An Adversarial Attack, where a Man is recognized as a woman, with very distinct features [Reference 25]

For example, Mahmood Sharif, et al. showed, in their paper "Accessorize to a Crime: Real and Stealthy Attacks on State-of-the-Art Face Recognition", that they fooled a Deep Neural Network trained for face verification, to identify a man (with distinct features) as a verified person, when in reality, the only verified person was a woman with different skin colour (as seen in Figure 2). Moreover, they were able to design a system to impersonate any person with just a single photograph of them, which is easily obtainable from the internet. This has serious implications in places where Face Verification is commonly used, such as China, where many restaurants and stores now support payments through automated Face Verification systems. Using the system shown in the paper above, an attacker could potentially impersonate any other citizen to purchase products in their name, leading to large scale financial frauds

Another example is that of safe transportation. Researchers at the Keen Security Lab have shown that they were able to fool a Tesla self-driving car to accelerate to 45 miles per hour despite seeing a stop sign, with simple vandalism like additions (as seen in Fig 3). This could be a major safety hazard once self-driving cars are available commercially. Such attacks could lead to more frequent crashes and even fatal injuries.

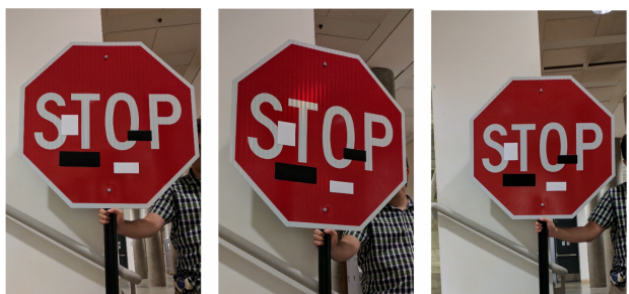

Fig 3 - An Adversarial Stop Sign: Recognized by the AI as a 45 Miles per hour sign[Reference 7]

There are numerous other places where such attacks could bear a huge cost to society. For example, Deep Neural Networks are now being commonly used to help identify skin cancer. An image of benign skin cancer can easily be turned into a malignant one with minimal noise addition and vice versa (as seen in Fig 4). With the lack of transparency in some hospitals, this could lead to numerous false diagnoses.

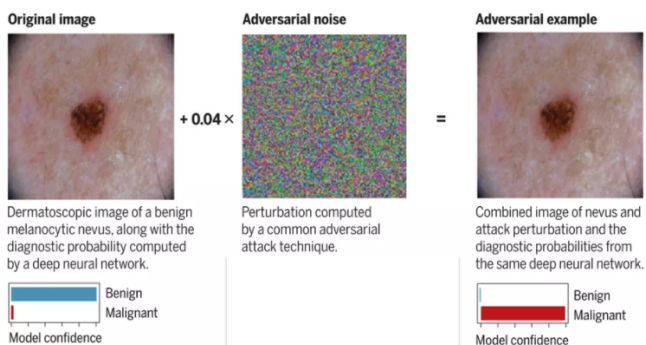

Fig 4 - A Benign Skin Cancer image that can be turned into a Malignant one [Reference 24]

## 1.2 Problem Statement

Hence, the need arises for simple, effective and robust techniques to make Deep Neural Network immune to such attacks. In this paper, we first implement two major adversarial attack techniques and measure the decline in accuracy of a baseline Neural Network against both techniques. We then explore some of the existing defense mechanisms and gauge their effectiveness of such defenses against a variety of factors and gauge their effectiveness and compare the improvement in accuracy. Finally, we conclude on which combination of defenses works best.

## 2. Literature Review

### 2.1 Related Papers

In the original paper on Adversarial attacks (Ian Goodfellow, et al.), the authors showed that Deep Neural Networks (DNNs) are highly sensitive to the changes in the input, meaning that the model's confidence in a prediction changed very quickly as the input image was changed. They then narrowed the perturbation space to see if the DNN was still highly sensitive to small changes to even a few pixels in the image. This observation led them to try and fool the model by adding small amounts of noise to the input. They aimed to add a small enough amount of noise for the image to look unaltered or easily recognizable as a human, yet fool a Deep Neural Network. They were successful in their attempt, showing the weakness of such Networks.

Discovering such attacks, Ian Goodfellow, et al. defined adversarial examples as follows: *"Adversarial examples are inputs to machine learning models that an attacker intentionally designed to cause the model to make mistakes"*.

Similar results were also produced for Convolutional Neural Networks on image tasks. This method, called the Fast Gradient Sign Method (described in detail later), can very quickly generate effective adversarial examples. As a prevention technique, researchers tried retraining the Deep Neural Network on data that included adversarial examples (known as adversarial training), and they were able to increase the performance of the network. However, subsequent research showed that small modifications such as increased noise levels and unique base images rendered the adversarial training technique futile.

However, until now, all attacks were untargeted, meaning that the misidentified class could not be chosen ex-ante by the attacker. Further research showed that by iteratively changing the input image, researchers were able to fool a model to believe any input image belonged to a particular class of their choice. This type of attack is known as a targeted attack and is described later on in the paper. For a more general approach to counter adversarial attacks, denoising was introduced, where pixels with a value more than a certain threshold were set to the brightest value, and those without were removed. This approach worked for baseline systems. However, with large increases in noise, the denoising approach didn't work, as the threshold couldn't be tuned in real-time. Moreover, such an approach is not feasible in larger datasets such as ImageNet, where the intensity of the pixel is important in the task.

Further research (Madry et al 2017), was aimed at training Deep Neural Networks to be more resilient to attacks, by choosing appropriate training data and including adversarial examples from a specific type of attack.

# 3. Implementation Strategy

## 3.1 Dataset - MNIST

I have used the standard Hand Digit Recognition MNIST dataset to conduct all my experiments. This dataset contains 60000 train and 10000 test images of various handwritten digits, from 0 to 9. The images are grayscale, having a dimension of 28 by 28 pixels. This dataset is one of the baseline datasets on which a lot of existing research has been formulated. Also, the Deep Neural Networks trained on this dataset are manageable on a personal computer, so it will be easier to run multiple experiments quickly. Finally, since Deep Neural Networks can easily achieve high performance on this dataset, I will be able to focus more on techniques to prevent adversarial attacks - which is the primary focus of this work - rather than developing a good baseline neural network for digit recognition.

## 3.2 Evaluation Methodology

The evaluation methodology is based on three factors, the accuracy of Deep Neural Networks, robustness to higher noise levels, and the technique used to generate the adversarial examples.

More concretely, for the MNIST dataset, I use the NN's accuracy to compare different techniques, since this is the most commonly used metric for benchmarking on this dataset. This is also an appropriate metric since all digits have almost equal representation in the train and test set, so any artificially high accuracy due to class imbalance is avoided.

For measuring the robustness of the Deep Neural Networks against noise, I evaluate the NN's accuracy against different increasing values of ε for the FGSM attack method. However, extremely high values of ε are avoided, since the images would then become pure noise rather than adversarial examples. This will help us to evaluate which Neural Networks are the most robust since many techniques only work for very low or high values of ε

Another important aspect to a good defense against adversarial attacks is their capability to defend against various kinds of attacks since the attacker has a wide array of attack techniques to choose from. Hence, for NN's that score well against larger values of ε in the FGSM technique, I evaluate their accuracy against the PGD technique, to see if they are still robust across attack techniques, which will help increase confidence in the durability of the defense.

## 4. Baseline Set-up

For evaluation of different attacks on a Neural Network, we created a baseline Neural Network trained only on the standard MNIST database. Our model had 3 convolutional layers, followed by a Max-Pooling layer. Finally, we had a Dropout layer and 3 other Dense Layers (Neural Network Graph in Appendices). The Neural Network was trained for 20 epochs on the standard dataset by which the validation accuracy peaked at 98%.

Accuracy of a baseline classifier against attack methods:

| Attack Method | Baseline Model Accuracy | Accuracy Reduction Due to Attack |
|---|---|---|
| FGSM | 23% | 75% |
| PGD | 20% | 78% |

Table 1 - Accuracy of Baseline Network against different attacks

As we can see, the same Neural Network that was able to classify images with a 98% accuracy now has an accuracy of around 23%, showing how susceptible Neural Networks are even for relatively simple tasks. While both attack methods reliably generate adversarial examples, we can see that the PGD attack technique is a stronger and more reliable way to generate adversarial examples since it fools the NN a higher percentage of the time.

## 5. Defenses, Results & Comparison

### 5.1 Adversarial Training

In the seminal paper on adversarial attacks by Ian Goodfellow, et al., researchers found that training the neural network on adversarial examples helps improve Neural Network robustness to some extent. However, new papers following this research showed even a small increase in the amount of noise (i.e an increase in the value of ε) would render this defense method useless. Since this is one of the most basic prevention techniques, it can act as a good benchmark to compare the effectiveness of techniques proposed in this paper. For training data, I converted all of the training data points into adversarial examples (creating 60,000 adversarial examples) with an ε of 0.1. To also make sure that our model doesn't forget how to classify normal genuine examples, we also included the entire training set, making a total of 120,000 images. When training the model, I found that Neural Networks would very quickly overfit to adversarial examples. Whereas a standard model could train on the MNIST dataset for 20 epochs without overfitting, when training on adversarial examples, within 2-3 epochs, the NN started overfitting to the adversarial examples we provided. Thus, if trained for longer, it did not perform well on adversarial examples generated from the test set. After training the NN for an appropriate number of epochs, we found that it had an accuracy of 61% against adversarial examples. Also, I found that increasing the noise by a small amount (from $\varepsilon = 0.1$ to $\varepsilon = 0.2$, the effectiveness of this technique went down drastically, to a level similar to the baseline Neural Network.

## 5.2 Denoising with a constant threshold

Another defense mechanism frequently mentioned is denoising images with a threshold[Reference 5]. For image-based tasks, this means that we choose a certain threshold $\alpha$ and if a pixel value is $> \alpha$, then it is set to be the maximum value (255), else if it is $< \alpha$ we set it to 0. This technique can effectively clean genuine (normal) images, as well as images with low noise (low values of ε), but not images with high amounts of noise. The reason for this is that the value $\alpha$ is not dynamic, meaning that it is a constant for a given model. Hence, $\alpha$ has to be tuned such that it allows enough number of pixels to pass through the filter to be able to make sense of the image (such as the pixels showing the 7 in Fig 5), but also not allow too many noisy pixels to pass through (such as the ones in the background). This is one of the main reasons behind the weakness of this defense. After experimenting with different thresholds, I arrived at $\alpha$ = 30% of 255 (meaning that a pixel with more than 76.5 intensity will be set to 255) as an appropriate threshold. After denoising with this threshold, the Neural Network was able to achieve 91% accuracy on the adversarial examples. However, upon increasing the value of $\varepsilon = 0.3$, the NN dropped to an accuracy of 5%, which is even worse than randomly guessing the digit (which would have a 10% accuracy). The reason for this is that with higher levels of noise, many more background pixels were able to pass through the filter and received a higher brightness, resulting in pronounced noise in "denoised" images (as shown in Fig 5). The same is true for the PGD attack, due to its ability to choose a few pixels and add high amounts of noise to those pixels (which can pass the filter).

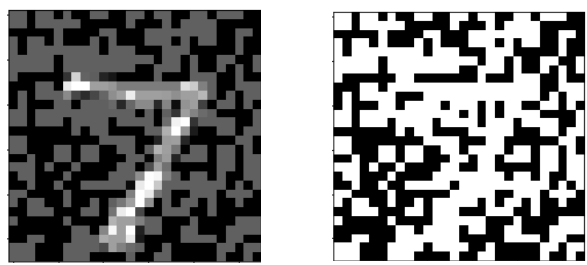

High noise Adversarial      Denoised Adversarial
Fig 5 - Effect of High Noise Adversarial on Denoising with a Constant Threshold

## 5.3 Neural Network to Flag Adversarial Examples

Adversarial examples are deliberately designed to fool a Neural Network, and so have malicious applications. If we can find a way to flag images as adversarial, we can remove them from the classification process, preventing Neural Networks from getting fooled. Such a system should have higher accuracy in test sets contaminated by adversarial examples.

Adversarial examples are very diverse in nature, even more than genuine images. Hence, an approach to learn to generalize to all kinds of adversarial examples would be to train another Deep Neural Network whose only objective is to detect and flag potentially adversarial examples. Thus, making a successful adversarial example now entails perturbing the image to circumvent two models, which effectively reduces the chance of success for any given adversarial example.

For this Neural Network (to flag adversarial examples), we used a smaller neural network because the task is much simpler in complexity compared to digit recognition. This network had 2 Convolutional layers, followed by a Dropout layer and 2 Dense layers (Neural Network Graph in the appendices).

This technique worked reasonably well for flagging adversarial examples generated using the Fast Gradient Sign Method. After excluding potential adversarial examples (as flagged by the Neural Network), the final classifier had an accuracy of 83%.

This technique also generalized to higher noise adversarial examples (for $\varepsilon = 0.2$ and 0.3). However, to test if this technique generalized to different attack methods, we tested its accuracy against the PGD attack. Here, the system only flagged 50% of the adversarial images correctly, making the final accuracy 60%.

Moreover, this suggests that such a flagging system may do even worse against other types of attacks (since it didn't even generalize to PGD, another gradient-based attack like FGSM, so it doesn't make a strong defense. These results are summarized in Table 2.

| Method | Baseline Accuracy | Flagging Adversarial Examples Accuracy |
|---|---|---|
| FGSM $(\varepsilon = 0.1)$ | 23% | 83% |
| PGD | 20% | 60% |

Table 2 - Results for Flagging Adversarial Examples

## 5.4 Batch Normalization at Test Time

Batch Normalization, as a technique, has been used to help Neural Networks generalize better to unseen images. With standard neural networks, each layer of neurons is dependent on the outputs of the previous layer of neurons. However, in the training process, all neurons update their weights with every batch to better understand the image. Hence, the inputs of later layers of neurons (same as the outputs of earlier layers of neurons) are different each time the neural network sees the same image during training, making it harder for these later neurons to update their weights accurately. Batch Normalization normalizes the outputs of each layer of neurons (same as the inputs of later neurons) in a neural network. Hence, following layers of neurons can learn faster.

In the current implementation of Batch Normalization, the mean and standard deviation used for normalization is calculated using only training data and then used for normalizing test data. Due to this, adversarial examples still have the capability to change the mean and standard deviation of the outputs of different layers of a Neural Network and have repercussions on the outputs of the final layer.

One potential approach to defend against adversarial attacks would be to calculate the mean and standard deviation used in normalization for every image that the neural network sees (irrespective of whether it is a train or test image). Hence, any change in inputs caused by adversarial examples will be suppressed, since they will be re-normalized and will have a smaller impact on later layers.

For this approach, we trained a fresh neural network with similar architecture as the baseline, but with a Batch Normalization layer after every Convolutional or Dense layer. When run against adversarial examples developed with the FGSM method ($\varepsilon = 0.1$), this modified network had an accuracy of 69%, compared to the 23% accuracy of the baseline model. Even with higher levels of noise ($\varepsilon = 0.2$), the model still had a much higher accuracy, which shows the robustness of this defense method.

However, given the accuracy of 69% was still low on an absolute basis, this method was not able to provide the acceptable level of resilience even against a single attack method (FGSM). Hence, we didn't test on adversarial examples generated by the PGD attack.

| Model | FGSM Accuracy ($\varepsilon = 0.1$) |
|---|---|
| Baseline Model | 23% |
| Batch Normalization at Test Time | 69% |

Table 3 - Results for Batch Normalization at Test Time

## 5.5 Dropout at Test Time

Dropout is a commonly used regularization technique to prevent Neural Networks from overfitting to the training data. Dropout, as the name suggests, randomly drops a chosen percentage of the neurons from layer(s) during training. This way, the Neural Network architecture changes every time it sees a new image during training (as the combination of neurons present is different), so it has to learn to distribute information amongst more neurons rather than relying only on a handful of neurons (since these few neurons might disappear randomly). However, this is currently used only during training, so when making predictions, the entire Neural Network is preserved in original form.

This technique could even be used when making predictions (in other words, at test time) to help defend against adversarial examples. The reason it might work is that without dropout, it is likely that a Neural Network will rely only on a few neurons, which in turn rely only on a few pixels of the image, to make predictions. Adversarial examples would change these few pixels and would be able to disrupt these neurons' predictions and hence the entire NN's predictions.

However, by dropping neurons, we use a different set of pixels of an image every time we make a prediction, so it becomes harder to generate adversarial examples. This is because it is impossible to know which neurons will be dropped in advance (and hence which set of pixels will be used in the final classification).

The problem with dropout, however, is that it reduces the NN's standard accuracy a lot since it has fewer neurons and hence less information to make a decision. Thus, the amount of dropout (percentage of neurons dropped in each layer) needs to be tuned based on the trade-off of adversarial accuracy and standard accuracy.

To find the optimum dropout levels, I used a grid search method (a brute force approach that tries out all possible values within a specified range and picks the value with the best performance) and evaluated different levels of dropout. To retain a high level of performance, I chose to use dropout in 3 of the 5 layers of the neural network, because dropping neurons from more than 3 layers would have resulted in a poor baseline accuracy.

My optimization objective for determining the best dropout percentage level was accuracy. However, I split this accuracy into two parts: accuracy on standard test data and accuracy on adversarial test data (an adversarial version of the test images). Final accuracy was computed on a weighted average, where the weights could be tuned based on the application.

In this case, I chose an equal weight of 0.5 for both validation datasets.

After running the grid search (changing dropout for one layer at a time), I arrived at the following optimal dropout values for each layer:

| Layer | % of Neurons Dropped |
|---|---|
| Dropout Layer 1 | 45 |
| Dropout Layer 2 | 3 |
| Dropout Layer 3 | 7.5 |

The grid search results intuitively make sense since the middle layers of a neural network have a larger number of neurons as compared to later layers, and hence can afford a higher percentage of dropped neurons without losing valuable information. The table below summarizes the accuracy of the model trained with the above dropout values.

| Data | Accuracy with drop out at test time | Accuracy of Baseline Model |
|---|---|---|
| Standard Test Data | 96% | 98% |
| FGSM ($\varepsilon = 0.1$) | 79% | 23% |
| PGD ($\varepsilon = 0.1$ & 0.3) | 62.5% | 20% |

Table 4 - Adversarial Results for Dropout at Test Time

As seen in Table 4, dropout at test time gave a significantly higher accuracy of 79% against the FGSM method compared to the baseline accuracy of only 23%. Moreover, this technique performed well against higher values of $\varepsilon$ as well, although the accuracy was lower here. This technique even generalized well to the PGD attack method (even for higher noise levels, ε = 0.3) as it retained a high accuracy of 62.5% as compared to the baseline accuracy of 20%. This decline in accuracy across attack methods was expected since PGD is a stronger attack as depicted by a lower accuracy of the baseline network.

## 5.6 Denoising Autoencoders

Autoencoders are neural network architectures made up of two symmetric neural networks (as shown in figure 6).

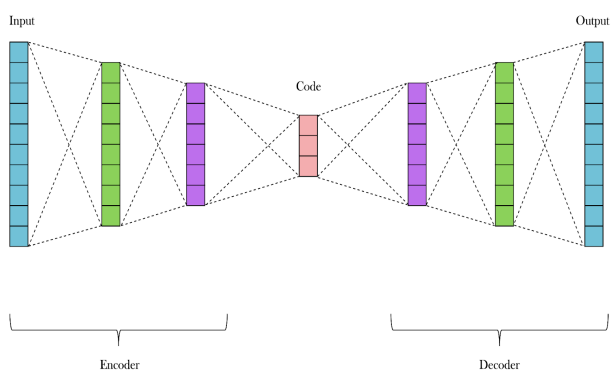

Fig 6 - General Structure of an Autoencoder
The first network is an encoder, which encodes a high dimensional input image into a small array of real numbers (of say size 10). The second network is a decoder which takes in this small array of real numbers and tries to reconstruct the entire original input image.

Autoencoders are one of the few categories of neural networks that are unsupervised by design, meaning that the label of the image is not required, only the images themselves are required. Hence, they can be used to perform general tasks for a wide variety of images. One application of autoencoders is denoising images. As shown by researchers, autoencoders can learn to clean relatively high noise images (non-adversarial) extremely well.

Adversarial examples are also, in a way, noisy images, where the noise is carefully calculated. A Denoising Autoencoder will not distinguish between artificial noise and adversarial noise and attempt to clean both similarly. Hence, a denoising autoencoder could potentially clean and remove noise from all images before feeding them to the standard neural network, thus boosting model accuracy.

We trained a denoising autoencoder to encode a 28 by 28 MNIST image into a 16-dimensional array, which was again decoded back to a 28 by 28 image after removing noise. Our autoencoder performed very well against images with randomly added noise, as shown in Fig 7.

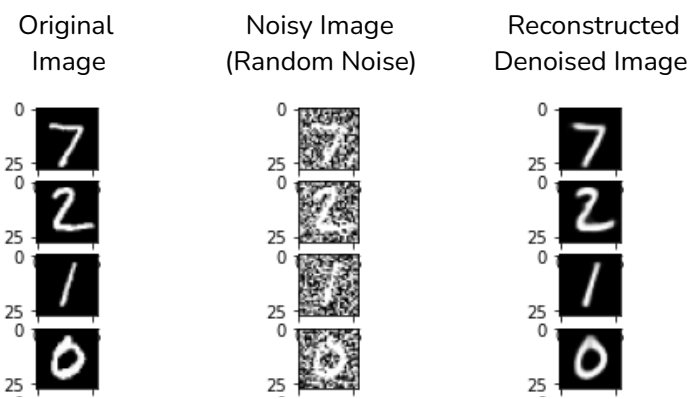

Fig 7 - Visual Results of Denoising Autoencoder on Random Noise

First, I trained the denoising autoencoder on images with artificially added random noise for 30 epochs, followed by training on adversarial examples for 2 epochs. To complete the pipeline, we first feed all input images (adversarial or not) through the denoising autoencoder to generate a denoised image. These denoised images were then fed to the baseline Neural Network for classification. After running this pipeline on an unseen test batch of adversarial examples, we observed an accuracy of 84%. Moreover, this pipeline generalized well for higher noise levels as well as the PGD attack, mainly because autoencoders can easily learn to remove general noise.

| Model | Standard Accuracy | FGSM Accuracy $\varepsilon = 0.1$ | PGD $\varepsilon = 0.1$ & $\varepsilon = 0.3$ |
|---|---|---|---|
| Standard Model | 98% | 23% | 20% |
| Denoising AutoEncoder + Standard Model | 95% | 84% | 65% |

Table 5 - Adversarial Results for using a Denoising Autoencoder to clean images

When investigating images where the entire network failed to classify adversarial examples correctly, I found that more often than not, the mistake could be attributed to the denoising autoencoder's failure to clean the images (example shown in Fig 8).

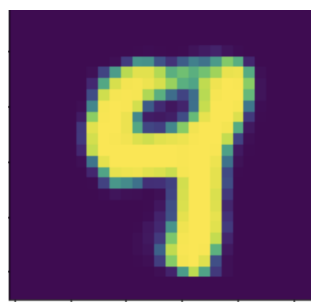

Predicted Label (by standard model) - 9
True Label - 4

Fig 8: A denoised adversarial image

However, we hypothesize that with an even larger training set of noisy and adversarial images, our autoencoder could reconstruct images with higher fidelity and improve the adversarial accuracy even beyond 84%. Moreover, advancements in autoencoders could also help generate better reconstructions of the noisy input image, which could further increase the effectiveness of this technique.

# 6. Discussion

Table 6 is a summary of the results of all the defense techniques proposed in this paper.

| Model | Standard Accuracy | Adversarial Accuracy | Generalizes Noise* | Generalizes Attack** |
|---|---|---|---|---|
| Baseline CNN | 98% | 23% | No | - |
| **Current Defenses** | | | | |
| Adversarial Training | 98% | 61% | No | - |
| Denoising with Constant Threshold | 98% | 91% | No | - |
| **Proposed Defenses** | | | | |
| Flagging Adversarial Examples | 97% | 83% | Yes | No |
| Batch Normalization (test time) | 97% | 69% | Yes | - |
| Dropout (test time) | 96% | 79% | Yes | Yes |
| Denoising AutoEncoder | 95% | 84% | Yes | Yes |

* More than 40% accuracy for $\varepsilon = 0.2$

** More than 50% accuracy for PGD attack

Table 6 - Summary of Results for All Defenses

## 6.1 Summary of Approaches and Results

From table 6, we can clearly see the need for strong, robust defenses against adversarial attacks, since a baseline model which had an accuracy of 98% on standard images, had an accuracy of only 23% when run against adversarial examples.

However, a good defense against adversarial examples should not only be able to have high adversarial accuracy, but also be able to generalize to higher noise levels. Else, attackers could simply increase the noise by a small amount and render the defense useless. Also, due to the wide variety of attack methods for attackers to choose from, a defense should be general enough to work on different attack methods.

The techniques currently used to defend against such attacks are adversarial training and denoising with a constant threshold. From the table, we can see that adversarial training had better accuracy, although not enough for practical purposes, and didn't generalize to higher noise levels. The reason for this is that adversarial training requires a variety of adversarial examples, but generating a wide enough sample is extremely difficult due to the variety of attack methods and noise levels available. Denoising with a constant threshold, on the other hand, had an extremely high adversarial accuracy. However, since the threshold for denoising is not dynamic, higher noise levels were able to pass through and reduce the accuracy to even worse than randomly guessing the digit. In such cases, the denoising defense was actually hurting the accuracy rather than helping.

We first experimented with the potential idea of flagging adversarial examples with another Neural Network. This approach had a high adversarial accuracy as it was accurately able to flag adversarial examples generated by the FGSM method. However, when moving onto different attack types (in this case PGD), it was only able to flag 50% of the adversarial examples, letting the other 50% pass through and reduce the model's accuracy. This may have been because this approach needs adversarial examples generated from a variety of attack methods, which is not practical in real-life applications.

We then tested the usage of a technique called Batch Normalization to be used during test time as well, instead of only during training. This approach, however, got an adversarial accuracy of 69%, which was relatively low compared to other methods. Even though it generalized to higher noise levels, the initial adversarial accuracy was too low to be used practically.

Finally, we see that both Dropout at test time and Denoising Autoencoders have a high adversarial accuracy (although Autoencoders are slightly better), and are also able to generalize to higher amounts of noise, as well as the type of attack. The difference between these two approaches, however, mainly lies in the computational complexity.

Autoencoders are much more computationally expensive since they are themselves made up of two smaller Neural Networks. For an image to be classified using the autoencoder pipeline, it has to first be encoded by the encoder, then decoded to be cleaned by the decoder, and then passed through the baseline model. On the other hand, Dropout at test time is negligibly more expensive than the standard baseline model and can run much faster than a Denoising Autoencoder.

## 6.2 Deciding on the Appropriate Defense

Based on these observations, the choice of defense depends upon the resources available for the application. Upon comparing the time difference between Autoencoder and the Dropout model for predicting labels for 10000 images, we found that the Autoencoder took 8x more time as compared to the Dropout model, clearly illustrating the difference in computational complexity.

Applications that permit bulkier Neural Networks, such as server-based operations, should make use of a denoising autoencoder, due to its higher accuracy and robustness (scaling up to other attack methods and noise). However, applications that involve edge ML (running Neural Networks on small processors such as phones or watches), should make use of the dropout at test time as a primary defense, because of its light resource requirement.

However, Dropout has an extremely strong regularization effect, so when used at test time, it reduces standard accuracy significantly. For a dataset like MNIST, it was possible to find a dropout level that maintained a substantial standard accuracy and improved adversarial accuracy as well. There might be applications where an extremely high standard accuracy is required, for example, biometrics, and here, dropout may reduce the standard accuracy to an unacceptable amount. Hence, in such applications, they may be forced to use the Denoising Autoencoder even if it is more computationally expensive.

## 6.3 Limitations and Future Work

Currently, in this paper, we have explored the robustness of defenses against only the FGSM and PGD attack methods. Although this gives some insight into which defenses would generalize, it is important to still evaluate these defenses against other attack methods, especially non-gradient based approaches.

Moreover, the MNIST dataset is a grayscale dataset. However, further experimentation needs to be done to validate that an effective autoencoder can be set up even for more complex image datasets, with special focus to RGB images.

Finally, we have also focused on untargeted attacks as part of the evaluation methodology discussed in this paper. Future work should be done to ensure that such defenses are resilient against targeted attacks as well, but we hypothesize that our results should still hold against targeted attacks.

## 7. Conclusion

This paper demonstrates the threat adversarial examples pose to commercially deployed Deep Neural Networks by fooling a high scoring Neural Network using two different attack methods (FGSM and PGD). To establish a baseline for comparing different defenses, the MNIST dataset was preferred due to large amounts of data availability as well as the relative simplicity of the task. This paper then also proposes a methodology for comparing defenses by using a variety of factors: adversarial accuracy, robustness to higher noise levels as well as the type of attack.

We then experimented with current defense mechanisms and evaluated their performance on our developed metric, and demonstrated that they are exploitable.

Upon comparison of different defense methods, using our metric, we found that Dropout at test time and Denoising Autoencoders were the most robust methodologies to safeguard against adversarial attacks and scaled well for higher noise levels as well as different types of attacks. Upon further comparison between methodologies, we concluded that for applications with easy access to computational power, a denoising autoencoder should be used, whereas for applications with limited access to computation power, the dropout at test time approach should be used.

# Appendix

## Source Code

https://github.com/Yushgoel/AdversarialAttack

## Neural Network Architectures

Baseline Neural Network:                    Neural Network to Flag Adversarial

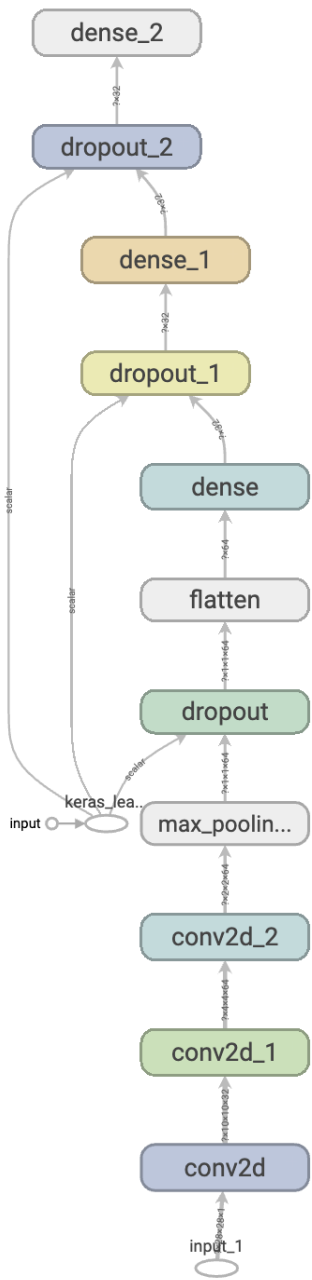
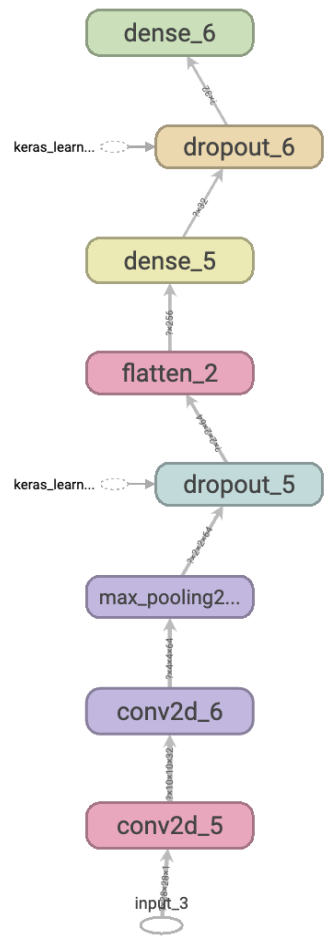

## Batch Normalization    |    Dropout at Test Time    |    Denoising Autoencoder
## at Test Time

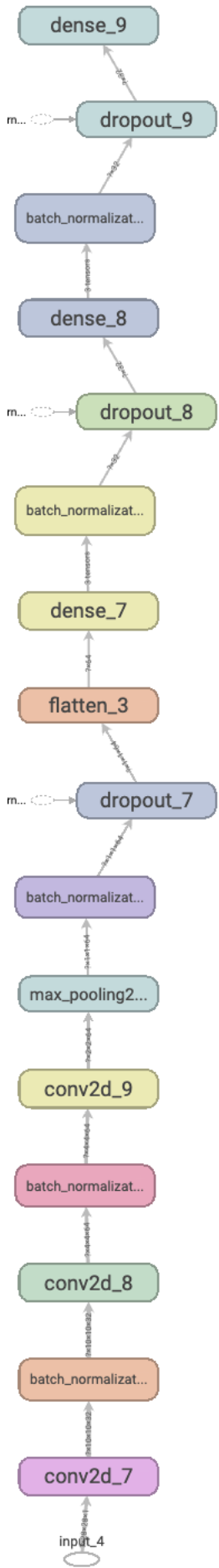

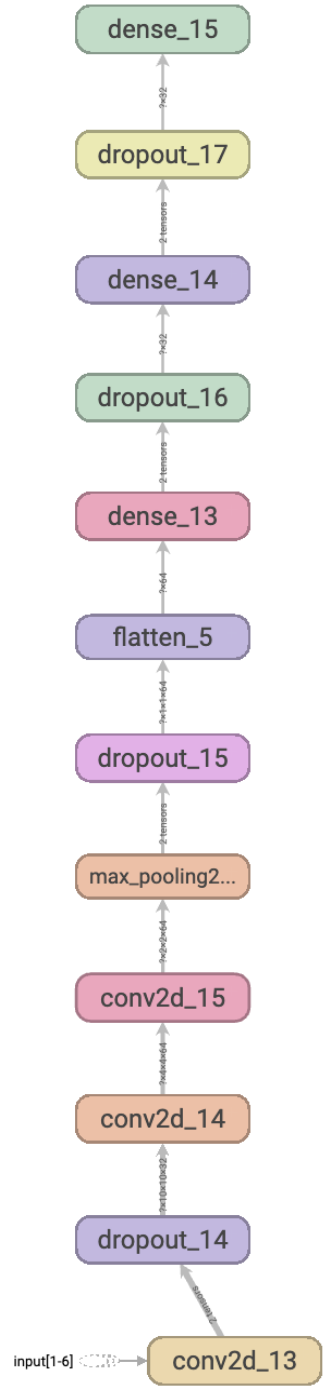

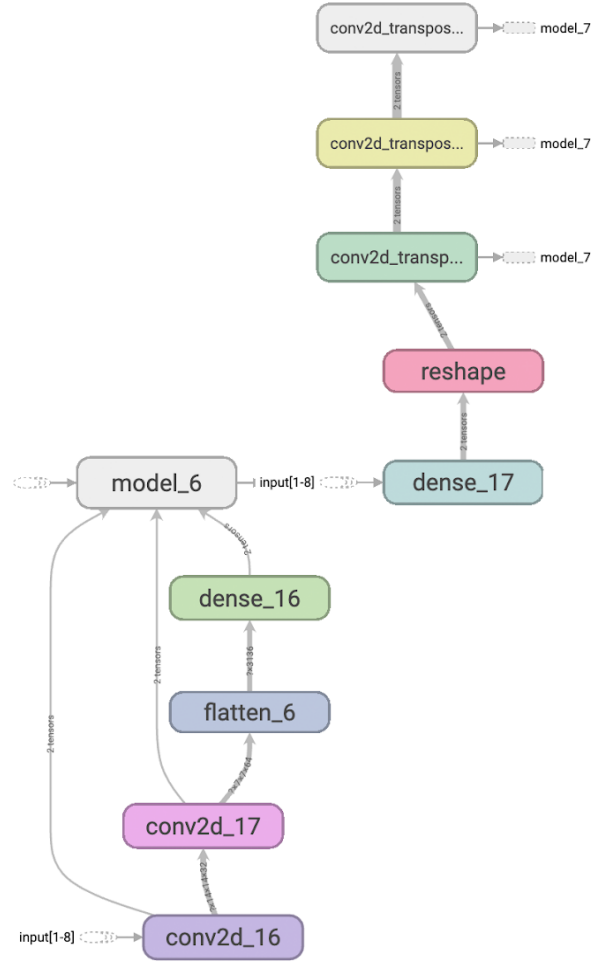